\documentclass[showpacs,10pt,twocolumn,prl]{revtex4-1}

\usepackage{amsmath}
\usepackage{amssymb}
\usepackage{graphicx}
\usepackage{amssymb}
\usepackage{graphics}
\usepackage{epsfig}
\usepackage{CJK}
\usepackage{color}
\usepackage{soul}

\begin{document}

\begin{CJK*}{GBK}{Song}
\title{Polaronic transport and thermoelectricity in Mn$_3$Si$_2$Te$_6$ single crystals}
\author{Yu Liu,$^{1,*}$ Zhixiang Hu,$^{1,2}$ Milinda Abeykoon,$^{3}$ Eli Stavitski,$^{3}$ Klaus Attenkofer,$^{3,\dag}$ Eric D. Bauer,$^{4}$ and C. Petrovic$^{1,2}$}
\affiliation{$^{1}$Condensed Matter Physics and Materials Science Department, Brookhaven National Laboratory, Upton, New York 11973, USA.\\
$^{2}$Materials Science and Chemical Engineering Department, Stony Brook University, Stony Brook, New York 11790, USA.\\
$^{3}$National Synchrotron Light Source II, Brookhaven National Laboratory, Upton, New York 11973, USA.\\
$^{4}$Los Alamos National Laboratory, Los Alamos, New Mexico 87545, USA.}
\date{\today}

\begin{abstract}
We carried out a comprehensive study of the structural, electrical transport, thermal and thermodynamic properties
in ferrimagnetic Mn$_3$Si$_2$Te$_6$ single crystals. Mn and Te $K$-edge X-ray absorption spectroscopy and synchrotron powder X-ray diffraction were measured to provide information on the local atomic environment and the average crystal structure. The dc and ac magnetic susceptibility measurements indicate a second-order paramagnetic to ferrimagnetic transition at $T_c$ $\sim$ 74 K, which is further confirmed by the specific heat measurement. Mn$_3$Si$_2$Te$_6$ exhibits semiconducting behavior along with a large negative magnetoresistance of -87\% at $T_c$ and relatively high value of thermopower up to $\sim$ 10 mV/K at 5 K. Besides the rapidly increasing resistivity $\rho(T)$ and thermopower $S(T)$ below 20 K, the large discrepancy between activation energy for resistivity $E_\rho$ and thermopower $E_S$ above 20 K indicates the polaronic transport mechanism. Furthermore, the thermal conductivity $\kappa(T)$ of Mn$_3$Si$_2$Te$_6$ is notably rather low, comparable to Cr$_2$Si$_2$Te$_6$, and is strongly suppressed in magnetic field across $T_c$, indicating the presence of strong spin-lattice coupling, also similar with Cr$_2$Si$_2$Te$_6$.
\end{abstract}
\maketitle
\end{CJK*}

\section{INTRODUCTION}

Layered transition-metal materials, an active area of research in condensed matter physics, have been extensively studied due to the exotic physical properties. Examples include the high temperature superconductivity in copper- and iron-based superconductors \cite{Bednorz,Wu,Kamihara}, the high thermoelectricity in cobaltites \cite{Terasaki,Masset}, the colossal magnetoresistance (MR) in manganites \cite{Kusters}, and the recently discovered long-range magnetic order in two-dimensional (2D) thin crystals in chromium or iron-based trichalcogenides and trihalides \cite{Huang,Gong}.

Cr$_2$(Si,Ge)$_2$Te$_6$ are intrinsic ferromagnetic (FM) semiconductors with peculiar Si-Si(Ge-Ge) dimers when compared with CrI$_3$ \cite{Casto,Zhang,Siberchicot,Carteaux,McGuire}. First-principles calculations predicted that the FM in Cr$_2$Si$_2$Te$_6$ survives even down to monolayer; FM in bilayer Cr$_2$Ge$_2$Te$_6$ and monolayer CrI$_3$ were experimentally observed \cite{Huang,Gong,Lebegue,Li,Lin}. The Cr atoms form a honeycomb lattice in the $\mathbf{ab}$ plane with Si or Ge in the center of hexagon and Cr is surrounded by octahedra of Te. In such a layered structure, there are various chemical bonds including intralayer Cr-Te ionic bonds, Si/Ge-Te covalent bonds, Si-Si/Ge-Ge metal bonds, and interlayer van der Waals (vdW) force. Then a low lattice thermal conductivity was observed in Cr$_2$(Si,Ge)$_2$Te$_6$ \cite{YangD,Tang,Tang1,Peng}. Interestingly, Cr$_2$Si$_2$Te$_6$ also features strong spin-phonon coupling; short-range in-plane FM correlations survive up to room temperature even though out-of-plane correlations disappear above 50 K \cite{Williams}.

Mn$_3$Si$_2$Te$_6$ is a little-studied three-dimensional (3D) analog of Cr$_2$Si$_2$Te$_6$ \cite{Vincent,Rimet,MAY}. Structurally the Mn$_2$Si$_2$Te$_6$ layer is composed of MnTe$_6$ octahedra that are edge sharing within the $\mathbf{ab}$ plane (Mn1 site) and along with Si-Si dimers [Fig. 1(a)], similar to Cr$_2$Si$_2$Te$_6$. Then the layers are connected by filling one-third of Mn atoms at the Mn2 site within interlayer, yielding a composition of Mn$_3$Si$_2$Te$_6$ \cite{MAY}. Recent neutron diffraction experiment shows that Mn$_3$Si$_2$Te$_6$ orders ferrimagnetically (FIM) below $T_c \approx 78$ K; the magnetic order has antiparallel alignment of Mn1 and Mn2 sublattices with an easy-plane anisotropy \cite{MAY}. The magnetization can be tuned by proton irradiation as a result of modification of Mn-Te-Mn exchange interactions \cite{Martinez,Sao}. By replacing Te by Se, the $T_c$ slightly decreases to 67 K as well as magnetization anisotropy {\color{blue}in} Mn$_3$Si$_2$Se$_6$ \cite{May}. A collinear FIM structure with the moments significantly tilting out of the $\mathbf{ab}$ plane was also confirmed in Mn$_3$Si$_2$Se$_6$, arising from shorter interatomic distances and modified local structure \cite{May}. Furthermore, a large MR was usually expected in some manganites, such as La$_{1-x}$Ca$_x$MnO$_3$ and Tl$_2$Mn$_2$O$_7$ \cite{MB,YT,PRL}, which has not been explored for Mn$_3$Si$_2$Te$_6$.

In this work we studied the structural, electrical and thermal transport properties of Mn$_3$Si$_2$Te$_6$ single crystals. Mn$_3$Si$_2$Te$_6$ exhibits a polaronic-type semiconducting behavior above 20 K along with a large negative MR $\sim$ -87\% at $T_c$ = 74 K. Both in-plane resistivity $\rho(T)$ and thermopower $S(T)$ increase rapidly below 20 K, and reach relatively high values of $\sim$ 10 $\Omega$ m and 10 mV K$^{-1}$, respectively, at 5 K. In-plane thermal conductivity $\kappa(T)$ is very low, comparable to observed in nanostructured or complex high-performance thermoelectric materials and is further suppressed in magnetic field across $T_c$ \cite{Vineis,Snyder,Beekman,ZhaoL,TakabatakeT}. This indicates strong spin-lattice coupling and potential for high thermoelectric performance.

\section{EXPERIMENTAL DETAILS}

Single crystals of Mn$_3$Si$_2$Te$_6$ were grown by a self-flux method as described in ref.\cite{YuLiu}. X-ray absorption spectroscopy was measured at 8-ID beamline of the National Synchrotron Light Source II (NSLS II) at Brookhaven National Laboratory (BNL) in the fluorescence mode. X-ray absorption near edge structure (XANES) and extended X-ray absorption fine structure (EXAFS) spectra were processed using the Athena software package. The EXAFS signal, $\chi(k)$, was weighed by $k^2$ to emphasize the high-energy oscillation and then Fourier-transformed in $k$ range from 2 to 10 {\AA}$^{-1}$ to analyze the data in $R$ space. Synchrotron X-ray diffraction (XRD) measurement was carried out in capillary transmission geometry using a Perkin Elmer amorphous silicon area detector placed 1000 mm downstream of the sample at 28-ID-1 (PDF) beamline of the NSLS II at BNL. The setup utilized a $\sim$ 74 keV ($\lambda$ = 0.1665 {\AA}) X-ray beam. 2D diffraction data were integrated using pyFAI software package \cite{PyFAI1,PyFAI2}. The Rietveld analysis was carried out using GSAS-II software package \cite{Toby}.

The magnetization data as a function of temperature and field were collected using a Quantum Design MPMS-XL5 system. The electrical resistivity, thermopower, and thermal conductivity were measured on a Quantum Design PPMS-9 with standard four-probe technique. Continuous measuring mode was used. The maximum heater power and period were set as 50 mW and 1430 s along with the maximum temperature rise of 3$\%$. The sample dimensions were measured by an optical microscope Nikon SMZ-800 with 10 $\mu$m resolution. The specific heat was measured on warming procedure between 1.95 and 300 K by the heat pulse relaxation method using a Quantum Design PPMS-9 with crystal mass $\sim$ 5 mg.

\begin{figure}
\centerline{\includegraphics[scale=1]{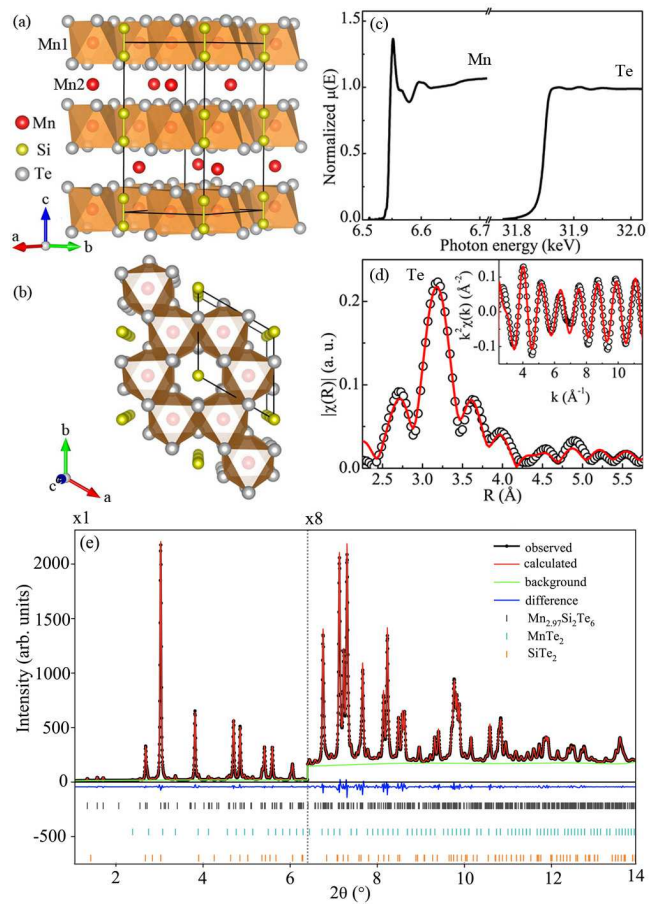}}
\caption{(Color online) Crystal structure of Mn$_3$Si$_2$Te$_6$ (space group: $P\bar{3}1c$) shown from the (a) side view and (b) top view, respectively. (c) Normalized Mn and Te $K$-edge X-ray absorption near edge structure (XANES) spectra. (d) Fourier transform magnitudes of the extended X-ray absorption fine structure (EXAFS) oscillations (symbols) for Te $K$-edge with the phase shifts correction. The model fits are shown as solid lines. Insets show the corresponding filtered EXAFS (symbols) with $k$-space model fits (solid lines). (e) Refinement of synchrotron powder XRD data of Mn$_3$Si$_2$Te$_6$ at room temperature, confirming main phase of Mn$_{2.97}$Si$_2$Te$_6$ (93 \%) with tiny impurities SiTe$_2$ (4 \%) and MnTe$_2$ (3 \%).}
\label{XRD}
\end{figure}

\begin{table}
\caption{\label{tab}Local bond distances extracted from the Te $K$-edge EXAFS spectra fits with fixed CN for Mn$_3$Si$_2$Te$_6$. CN is the coordination number based on crystallographic value, R is the interatomic distance, and $\sigma^2$ is the Debye Waller factor.}
\begin{ruledtabular}
\begin{tabular}{lllll}
   distance & CN & R ({\AA}) & $\Delta$R ({\AA}) & $\sigma^2$ ({\AA}$^2$)\\
  \hline
  Te-Si.1 & 1 & 2.50 & 0.32 & 0.026 \\
  Te-Mn2.1 & 2 & 2.92 & 0.09 & 0.003 \\
  Te-Mn1.2 & 1 & 2.95 & 0.09 & 0.003 \\
  Te-Si.2 & 1 & 3.84 & 0.14 & 0.019 \\
  Te-Te.1 & 2 & 3.97 & 0.14 & 0.019 \\
  Te-Te.2 & 2 & 4.03 & 0.14 & 0.019 \\
\end{tabular}
\end{ruledtabular}
\end{table}

\begin{table}
\caption{\label{tab}Structural parameters for Mn$_3$Si$_2$Te$_6$ obtained from the synchrotron powder XRD at room temperature.}
\begin{ruledtabular}
\begin{tabular}{llllll}
  \multicolumn{1}{c}{Chemical:} &\multicolumn{2}{c}{Mn$_{2.97}$Si$_2$Te$_6$} &\multicolumn{2}{c}{Space group:} &\multicolumn{1}{c}{$P\bar{3}1c$}\\
  \multicolumn{1}{c}{$a$ ({\AA})} &\multicolumn{2}{c}{7.0346(3)} &\multicolumn{2}{c}{$\alpha$ ($^\circ$)} &\multicolumn{1}{c}{60}\\
  \multicolumn{1}{c}{$b$ ({\AA})} &\multicolumn{2}{c}{7.0346(3)} &\multicolumn{2}{c}{$\beta$ ($^\circ$)} &\multicolumn{1}{c}{60}\\
  \multicolumn{1}{c}{$c$ ({\AA})} &\multicolumn{2}{c}{14.2435(4)} &\multicolumn{2}{c}{$\gamma$ ($^\circ$)} &\multicolumn{1}{c}{120}\\
  \multicolumn{1}{c}{$V$ ({\AA}$^3$)} &\multicolumn{2}{c}{610.41(3)} &\multicolumn{2}{c}{Density (g/cm$^3$)} &\multicolumn{1}{c}{5.3576}\\
  \hline
   Site & $x$ & $y$ & $z$ & Occ. & U$_{iso}$ ({\AA}$^2$) \\
  \hline
   Te & 0.3407(5) & 0.0080(7) & 0.1280(1) & 1 & 0.0112(2) \\
   Mn1 & 0.33333 & 0.66667 & -0.0031(2) & 1 & 0.0040(8) \\
   Si & 0 & 0 & 0.0775(6) & 1 & 0.004(2) \\
   Mn2 & 0.33333 & 0.66667 & 0.25 & 0.966 & 0.004(2)\\
  \hline
  Distance &  & R ({\AA}) & Distance &  & R ({\AA})\\
  \hline
  Si-Si & $\times$1 & 2.21(1) & Si-Te & $\times$3 & 2.476(5) \\
  Mn1-Te & $\times$3 & 2.901(4) & Mn2-Te & $\times$6 & 2.943(3)\\
  Mn1-Te & $\times$3 & 3.022(4) & Mn1-Mn2 & $\times$2 & 3.604(2)\\
  Si-Te & $\times$3 & 3.766(7) & Te-Te & $\times$2 & 3.968(5)\\
\end{tabular}
\end{ruledtabular}
\end{table}

\section{RESULTS AND DISCUSSIONS}

Figure 1(a,b) shows the crystal structure of Mn$_3$Si$_2$Te$_6$ from the side and the top views, respectively. The Mn1 atoms form a honeycomb arrangement of the edge-shared MnTe$_6$ octahedra. The Si pairs form Si$_2$Te$_6$ ethane-like groups. Then the Mn$_2$Si$_2$Te$_6$ layers are connected by the interlayer Mn2 atoms; the Mn2 atoms form a triangular lattice. Figure 1(c) exhibits the normalized Mn and Te $K$-edge XANES spectra. The threshold energies ($E_0$) obtained from the peak of derivative curve are about 6.544 and 31.849 keV for Mn and Te, respectively, indicating a mixed valence but close to the Mn$^{2+}$ state \cite{Subias}. Figure 1(d) shows the Fourier transform magnitudes of EXAFS spectra of Te. In a single-scattering approximation, the EXAFS can be described by \cite{Prins}:
\begin{align*}
\chi(k) = \sum_i\frac{N_iS_0^2}{kR_i^2}f_i(k,R_i)e^{-\frac{2R_i}{\lambda}}e^{-2k^2\sigma_i^2}sin[2kR_i+\delta_i(k)],
\end{align*}
where $N_i$ is the number of neighbouring atoms at a distance $R_i$ from the photoabsorbing atom. $S_0^2$ is the passive electrons reduction factor, $f_i(k, R_i)$ is the backscattering amplitude, $\lambda$ is the photoelectron mean free path, $\delta_i$ is the phase shift, and $\sigma_i^2$ is the correlated Debye-Waller factor measuring the mean square relative displacement of the photoabsorber-backscatter pairs. The main peak has been phase shift corrected by the standard for the nearest-neighbor Te-Si [2.50(32) {\AA}] in the Fourier transform magnitudes of EXAFS [Fig. 1(d)], corresponding to two different Te-Mn2 and Te-Mn1 bond distances with 2.92(9) {\AA} and 2.95(9) {\AA}, respectively, extracted from the model fit with fixed coordination number and $\sigma^2 = 0.003$ {\AA}$^2$. It matches well within experimental errors with the average crystal structure analysis \cite{Vincent}. Figure 1(e) shows the refinement result of synchrotron powder XRD data of Mn$_3$Si$_2$Te$_6$ at room temperature (space group $P\overline{3}1c$). The determined lattice parameters are $a$ = 7.0346(3) {\AA} and $c$ = 14.2435(4) {\AA}. The detailed structural parameters for Mn$_3$Si$_2$Te$_6$ local and average structure are summarized in Tables I and II, respectively. No significant difference is observed, suggesting that average crystallographic methods well describe local structure units.

\begin{figure}
\centerline{\includegraphics[scale=1]{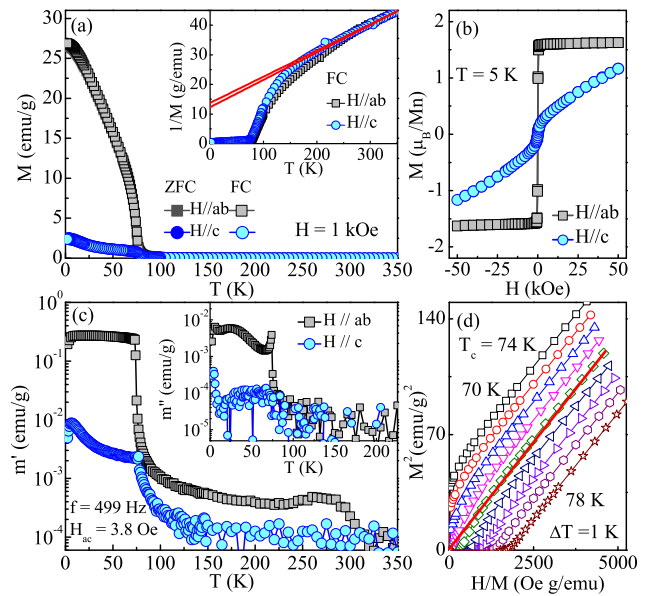}}
\caption{(Color online) (a) Temperature dependence of zero field cooling (ZFC) and field cooling (FC) dc magnetization $M(T)$ measured at $H$ = 1 kOe with both $\mathbf{H\parallel ab}$ and $\mathbf{H\parallel c}$ for Mn$_3$Si$_2$Te$_6$ single crystal. Inset shows the FC $1/M(T)$ vs $T$ curves fitted by the Curie-Weiss law from 250 to 350 K. (b) The isothermal magnetization $M(H)$ measured at $T$ = 5 K for both directions. (c) Temperature dependence of ac susceptibility real part $m^\prime(T)$ and imaginary part $m^{\prime\prime}(T)$ (inset) measured in ac field of 3.8 Oe and with frequency of 499 Hz. (d) The Arrott plot $M^2$ vs $H/M$ with $\mathbf{H\parallel ab}$.}
\label{MTH}
\end{figure}

\begin{figure}
\centerline{\includegraphics[scale=1]{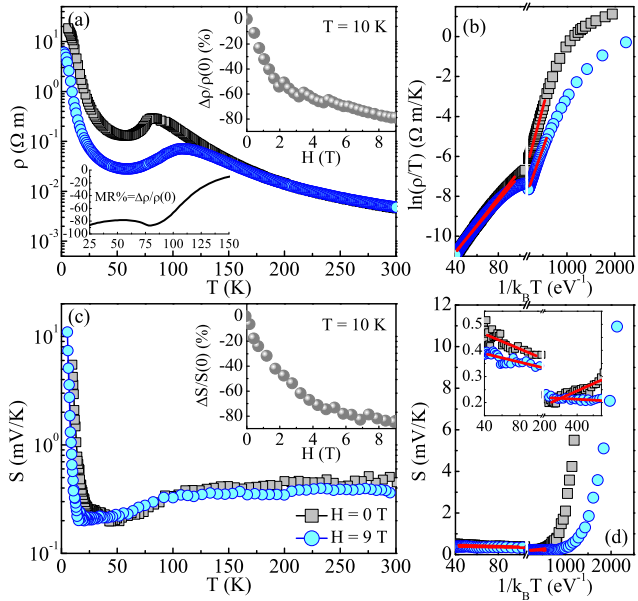}}
\caption{(Color online) (a) Temperature dependence of in-plane resistivity $\rho(T)$ of Mn$_3$Si$_2$Te$_6$ single crystal in 0 and 9 T. Inset shows the field-dependent $\rho(T)$ at 10 K. (b) The ln($\rho/T$) vs $1/k_BT$ curve fitted by adiabatic small polaron hopping model $\rho(T) = ATexp(E_\rho/k_BT)$ with $E_\rho$ being the activation energy. (c) Temperature dependence of in-plane thermopower $S(T)$ of Mn$_3$Si$_2$Te$_6$ single crystal in 0 and 9 T. Inset shows the field-dependent $S(T)$ at 10 K. (d) The $S(T)$ vs $1/k_BT$ curve fitted by the formula $S(T) = (k_B/e)(\alpha+E_S/k_BT)$ with $E_S$ being the activation energy.}
\label{3}
\end{figure}

Figure 2(a) shows the temperature dependence of magnetization measured in $H$ = 1 kOe applied in the $\mathbf{ab}$ plane and along the $\mathbf{c}$ axis, respectively. A sharp increase in $M(T)$ is observed upon cooling below $T_c$ for $\mathbf{H\parallel ab}$, and the values of $M(T)$ are much larger than those for $\mathbf{H\parallel c}$, indicating an easy-plane anisotropy. The zero field cooling (ZFC) and field cooling (FC) data for both orientations overlap well to each other below $T_c$, suggesting the absence of high-temperature ferromagnetic impurities. The $1/M$ vs $T$ data [inset in Fig. 2(a)] from 250 to 350 K can be fitted by the Curie-Weiss law, giving an effective moment of $\mu_{eff}$ = 5.33(9) $\mu_B$/Mn for $\mathbf{H\parallel ab}$ and 5.45(3) $\mu_B$/Mn for $\mathbf{H\parallel c}$, respectively, close to the expected value of Mn$^{2+}$. The derived Weiss temperatures are -133(6) K and -155(2) K, indicating that strong antiferromagnetic correlation exists in the paramagnetic phase. The increase of $M(T)$ below $T_c$ together with the large negative values of Weiss temperature is consistent with a ferrimagnetic ground state \cite{Rimet}. The antiparallel alignment of moments from Mn1 and Mn2 sublattices results in a FIM order with an easy-plane anisotropy \cite{MAY}. Isothermal magnetization at $T$ = 5 K [Fig. 2(b)] shows saturation moment of $M_s \approx$ 1.6 $\mu_B$/Mn for $\mathbf{H\parallel ab}$ and a small FM component for $\mathbf{H\parallel c}$, confirming the easy-plane anisotropy. {\color{blue}The remanent moment is negligible} for both orientations, {\color{blue}which is} different from the initial report \cite{Rimet}, further indicates the single crystals are of high quality. Figure 2(c) shows the temperature dependence of ZFC ac susceptibility measured with oscillated ac field of 3.8 Oe and frequency of 499 Hz, confirming the FIM transition and easy-plane anisotropy. The $T_c$ = 74 K can be determined by the Arrott plot of $M^2$ vs $H/M$ [Fig. 2(d)] \cite{Arrott1}, in which the line at $T_c$ passes through the origin. This is in agreement with previous reports \cite{Rimet,MAY,YuLiu}.

Figure 3(a) shows the temperature dependence of in-plane electrical resistivity $\rho(T)$ for Mn$_3$Si$_2$Te$_6$ single crystal. The $\rho(T)$ increases as the temperature is decreased from a value of 0.46 $\Omega$ cm at 300 K, exhibiting a typical semiconducting behaviour. There is a sharp dip in zero-field $\rho(T)$ around $T_c$, mostly arising from the suppression of spin fluctuations stemming from FIM order, which shifts to higher temperature in 9 T. The calculated MR [$=(\rho_{9T}-\rho_{0T})/\rho_{0T}\times100\%$] is plotted in the left inset, showing a maximum negative MR of $\sim$ -87\% at $T_c$. This large negative MR was also observed in the pyrochlore magnaites \cite{PRL}, in which the ultra-low-density carriers can form magnetic polarons dressed by mean-field ferromagnetic spin fluctuations in an intermediate temperature regime above $T_c$, and the large MR emerges via suppressing the spin fluctuations \cite{PRL}. For the electrical transport mechanism we consider the thermally activated model $\rho(T) = \rho_0 exp(E_\rho/k_BT)$ and the adiabatic small polaron hopping model $\rho(T) = AT exp(E_\rho/k_BT)$ where $k_B$ = 8.617 eV K$^{-1}$ is the Boltzmann constant and $E_\rho$ is activation energy. Figure 3(b) shows the fitting result of the adiabatic small polaron hopping model in two temperature ranges 300-120 K and 50-20 K. The extracted activation energies $E_\rho$ are 69(1) meV for 300-120 K and 8.8(1) meV for 50-20 K in 0 T, which are 64(1) meV and 6.8(1) meV in 9 T, respectively.

Since the $\rho(T)$ data can also be fitted by the thermally activated model, we then measured the thermopower $S(T)$ to distinguish {\color{blue}between} these two models. The $S(T)$ shows positive values in the whole temperature range with a relatively large value of 507 $\mu$V K$^{-1}$ at 300 K, as depicted in Fig. 3(c), indicating dominant hole-type carriers. Above 150 K, the weak temperature dependence of $S(T)$ is probably due to electronic diffusion with small change of carrier concentration at high temperatures. With decreasing temperature, the $S(T)$ decreases and changes its slope across $T_c$. In the same two temperature ranges it can be fitted with the equation $S(T) = (k_B/e)(\alpha+E_S/k_BT)$ \cite{Austin}, where $E_S$ is activation energy and $\alpha$ is a constant. The derived activation energies for thermopower $E_S$ = 17(2) meV for 300-120 K and 3.2(2) meV for 50-20 K in 0 T, which are 10(2) meV and 0.4(1) meV in 9 T, respectively [Fig. 3(d) and inset]. As is seen, the values of $E_S$ are much smaller than $E_\rho$. This typically reflects a polaron transport mechanism of carriers. According to the polaron model, the $E_S$ is the energy required to activate the hopping of carriers, while the $E_\rho$ is the sum of the energy needed for the creation of carriers and activating the hopping of carriers \cite{Austin}. Therefore, within the polaron hopping model the activation energy $E_S$ is smaller than $E_\rho$. It is of high interest to note that both $\rho(T)$ and $S(T)$ increase rapidly below 20 K, gradually deviating from the polaron-transport behavior, and reach relatively high values of $\sim$ 10 $\Omega$ m and 10 mV K$^{-1}$, respectively, at 5 K. Field-dependent $\rho(T)$ and $S(T)$ were collected at $T$ = 10 K [insets in Fig. 3(a,c)], showing similar large negative values of -87\% and -89\%, respectively, in field change of 9 T. This is consistent with dominant electronic mechanism of thermopower.

\begin{figure}
\centerline{\includegraphics[scale=1]{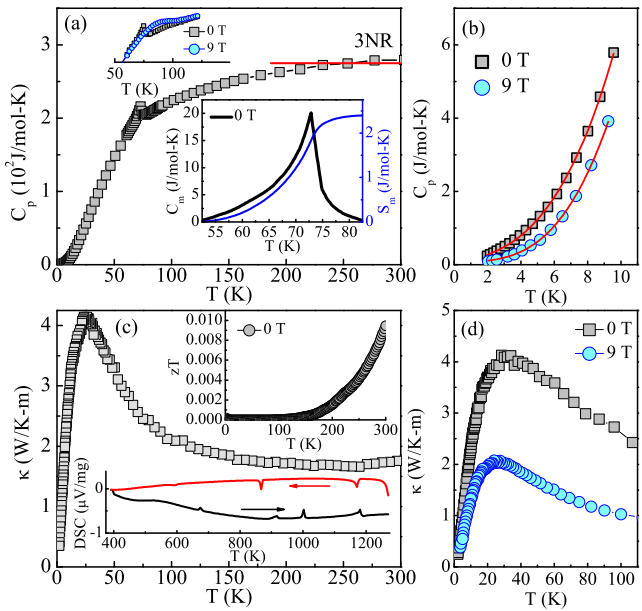}}
\caption{(Color online) Temperature dependence of (a) specific heat $C_p(T)$ and (b) the low temperature part fitted by using $C_p(T) = \gamma T+ \beta T^3 + \delta T^{3/2}$ for Mn$_3$Si$_2$Te$_6$ single crystal. Insets in (a) show the $C_p(T)$ near $T_c$ in 0 and 9 T, and the estimated magnetic contribution $C_m(T)$ (left axis) and derived magnetic entropy (right axis). (c) Temperature dependence of in-plane thermal conductivity $\kappa(T)$. Insets show the estimated zT = $S^2T/\kappa\rho$ and the differential scanning calorimetry (DSC) curve for Mn$_3$Si$_2$Te$_6$ single crystal. (d) The $\kappa(T)$ in 0 and 9 T at low temperatures.}
\label{MTH}
\end{figure}

Figure 4(a) shows the temperature dependence of specific heat $C_p(T)$ for Mn$_3$Si$_2$Te$_6$ single crystal. A clear $\lambda$-type peak in zero-field that corresponds to the second-order FIM transition was observed at $T_c$, which becomes broad and shifts to higher temperature in 9 T [top inset in Fig. 4(a)]. The magnetic entropy $S(T)$ = 2.4 J mol$^{-1}$ K$^{-1}$ is calculated from $S(T) = \int_0^T C_p(T,H)/TdT$ in temperature range from 55 to 85 K [bottom inset in Fig. 4(a)]. It is only 16\% of the value expected for ordering of $S$ = 5/2 local moments on Mn$^{2+}$ [$Rln(2S + 1)=14.9$ J mol$^{-1}$ K$^{-1}$], indicating short-range correlations or 2D magnetic order formation above $T_c$, which was previously observed in diffuse magnetic scattering experiment \cite{MAY}. The short-ranged, 2D magnetic correlations persist to very high temperature above $T_c$, and that much of missing entropy is released gradually as short-ranged order develops, which was also observed in Cr$_2$Si$_2$Te$_6$ \cite{Casto}. The low temperature $C_p(T)$ from 2 to 10 K {\color{blue}may be fit} by $C_p(T) = \gamma T+ \beta T^3 + \delta T^{3/2}$, where $\gamma T$ is the Sommerfeld electronic part, $\beta T^3$ is low-temperature limit of Debye phonon part, and $\delta T^{3/2}$ is low-temperature approximation of spin wave contribution, respectively \cite{Gopal}. The derived $\gamma$ and $\beta$ are 114(2) mJ mol$^{-1}$ K$^{-2}$ and 5.35(3) mJ mol$^{-1}$ K$^{-4}$ for $H$ = 0 T, and 115(15) mJ mol$^{-1}$ K$^{-2}$ and 5.7(1) mJ mol$^{-1}$ K$^{-4}$ for $H$ = 9 T, respectively, while $\delta$ is negligible. The large value of $\gamma$ in magnetic semiconductor could point to a constant density of states of magnetic excitations arising from high-temperature short range correlations like spin-glass state \cite{MAY,HuangC}. The Debye temperature $\Theta_D$ = 159(1) K for $H$ = 0 T and 155(1) K for $H$ = 9 T, respectively, can be calculated from $\beta$ using $\Theta_D = (12\pi^4NR/5\beta)^{1/3}$, where $N$ is the number of atoms per formula unit and $R$ = 8.314 J mol$^{-1}$ K$^{-1}$.

Figure 4(c) exhibits the temperature dependence of in-plane thermal conductivity $\kappa(T)$ of Mn$_3$Si$_2$Te$_6$ single crystal. In general, the $\kappa(T)$ consists of the electronic part $\kappa_e$ and the lattice phonon term $\kappa_{L}$, i. e., $\kappa = \kappa_e + \kappa_{L}$. The $\kappa_e$ part can be estimated from the Wiedemann-Franz law $\kappa_e = L_0T/\rho$ with $L_0$ = 2.45 $\times$ 10$^{-8}$ W $\Omega$ K$^{-2}$ and $\rho$ is the measured electrical resistivity. The calculated $\kappa_e$ is smaller than 0.003 W K$^{-1}$ m$^{-1}$ due to its large resistivity, indicating predominant lattice conductivity. The features of $\kappa(T)$ are similar with Cr$_2$Si$_2$Te$_6$ single crystal; a relatively low value of 1.76 W K$^{-1}$ m$^{-1}$ at 300 K and a typical phonon peak of 4.14 W K$^{-1}$ m$^{-1}$ around 26 K, rapidly increasing across $T_c$, almost ten times smaller than that of Mn$_3$Si$_2$Se$_6$ \cite{May}. The low thermal conductivity in Mn$_3$Si$_2$Te$_6$ might be contributed by the multiple types of chemical bonds like in Cr$_2$Si$_2$Te$_6$ as well as the random distribution of tiny impurities MnTe$_2$ and SiTe$_2$. Considering the low $\kappa(T)$ and high $S(T)$, we further estimated the temperature dependence of figure of merit (zT) for Mn$_3$Si$_2$Te$_6$ single crystal [inset in Fig. 4(c)], which increases with increasing temperature and reaches $\sim$ 0.01 at 300 K. A higher zT value should be observed at higher temperatures and will possibly enhanced by effective band engineering and/or charge doping. The $\kappa(T)$ is strongly suppressed in a magnetic field of 9 T at low temperatures [Fig. 4(d)], indicating similar strong spin-coupling in Mn$_3$Si$_2$Te$_6$, comparable to Cr$_2$Si$_2$Te$_6$. The $\kappa(T)$ in Mn$_3$Si$_2$Se$_6$ is also strongly field-dependent but with opposite trend \cite{May}, which requires further in-depth theoretical study. Yet, we note that not only lattice vibrations are involved in heat conduction, both below and above the magnetic transition. $\kappa$($T$) is insensitive to the onset of ferrimagnetic order whereas it is strongly suppressed in 9 T magnetic field [Fig. 4(d)]. The suppression points to close relation of the heat conduction and magnetic excitations that develop above the $T_c$. This agrees with small entropy fraction released below temperature of magnetic order and the presence of short-range correlations or 2D magnetic order formation at high temperatures [Fig. 4(a) inset].

Layered vdW magnetic materials show a great promise for thermoelectric applications, mostly due to rather low thermal conductivity which arises due to mixed bonding character which promotes phonon anharmonicity and scattering at the interfaces between vdW layers, low sound velocity, possible point defects and tunable electronic conduction \cite{YangD,Tang}. Polaronic transport, consistent with an enhanced electronic specific heat [Fig. 4(b)], could also contribute to relatively large thermopower in Mn$_3$Si$_2$Te$_6$ due to the large density of states around the energy gap edges.

\section{CONCLUSIONS}

In summary, Mn$_3$Si$_2$Te$_6$ features polaronic transport and very low values of thermal conductivity which is further suppressed in magnetic field. Thermoelectric figure of merit zT features relatively small value of 0.01 around 300 K, but relatively constant thermopower, and strongly decreasing trend of electrical resistivity and thermal conductivity suggest higher values at higher temperatures that could be further enhanced by chemical substitution or defect engineering.

\textit{Note added}. We recently became aware that Y. Ni \emph{et al.} also reported the transport properties of Mn$_3$Si$_2$Te$_6$ \cite{Ni}, in agreement with our results.

\section*{Acknowledgements}

Work at BNL is supported by the Office of Basic Energy Sciences, Materials Sciences and Engineering Division, U.S. Department of Energy (DOE) under Contract No. DE-SC0012704. This research used the 8-ID and the 28-ID-1 beamline of the NSLS II, a U.S. DOE Office of Science User Facility operated for the DOE Office of Science by BNL under Contract No. DE-SC0012704. {\color{blue}Y.L. acknowledges a Director$^\prime$s Postdoctoral Fellowship through the Laboratory Directed Research and Development program of the Los Alamos National Laboratory. E.D.B. was supported by the Office of Basic Energy Sciences, Materials Sciences and Engineering Division, U.S. DOE under the ``Quantum Fluctuations in Narrow-Band Systems."}\\

$^{*}$Present address: Los Alamos National Laboratory, MS K764, Los Alamos NM 87545, USA.\\

$^{\dag}$Present address: ALBA Synchrotron Light Source, Cerdanyola del Valles, E-08290 Barcelona, Spain.


\begin{references}
\bibitem{Bednorz} J. G. Bednorz and K. A. Muller, Z Phys. B \textbf{64}, 189 (1986).
\bibitem{Wu} M. K. Wu, J. R. Ashburn, C. J. Torng, P. H. Hor, R. L. Meng, L. Gao, Z. J. Huang, Y. Q. Wang, and C. W. Chu, Phys Rev Lett \textbf{58}, 908 (1987).
\bibitem{Kamihara} Y. Kamihara, T. Watanabe, M. Hirano, and H. Hosono, J. Am. Chem. Soc. \textbf{130}, 3296 (2008).
\bibitem{Terasaki} I. Terasaki, Y. Sasago and K. Uchinokura, Phys. Rev. B \textbf{56}, R12685 (1997).
\bibitem{Masset} A. C. Masset, C. Michel, A. Maignan, M. Hervieu, O. Toulemonde, F. Studer, B. Raveau and J. Hejtmanek, Phys. Rev. B \textbf{62}, 166 (2000).
\bibitem{Kusters} R. M. Kusters, J. Singleton, D. A. Keen, R. McGreevy, and W. Hayes, Physica B: Condensed Matter \textbf{155}, 362 (1989).
\bibitem{Huang} B. Huang, G. Clark, E. Navarro-Moratalla, D. R. Klein, R. Cheng, K. L. Seyler, D. Zhong, E. Schmidgall, M. A. McGuire, D. H. Cobden, W. Yao, D. Xiao, P. Jarillo-Herrero, and X. D. Xu, Nature \textbf{546}, 270 (2017).
\bibitem{Gong} C. Gong, L. Li, Z. L. Li, H. W. Ji, A. Stern, Y. Xia, T. Cao, W. Bao, C. Z. Wang, Y. Wang, Z. Q. Qiu, R. J. Cava, S. G. Louie, J. Xia, and X. Zhang, Nature \textbf{546}, 265 (2017).
\bibitem{Casto} L. D. Casto, A. J. Clune, M. O. Yokosuk, J. L. Musfeldt, T. J. Williams, H. L. Zhuang, M. W. Lin, K. Xiao, R. G. Hennig, B. C. Sales, J. Q. Yan, and D. Mandrus, APL Mater. \textbf{3}, 041515 (2015).
\bibitem{Zhang} X. Zhang, Y. Zhao, Q. Song, S. Jia, J. Shi, and W. Han, Jpn. J. Appl. Phys. \textbf{55}, 033001 (2016).
\bibitem{Siberchicot} B. Siberchicot, S. Jobic, V. Carteaux, P. Gressier, and G. Ouvrard, Phys. J. Chem. \textbf{100}, 5863 (1996).
\bibitem{Carteaux} V. Carteaux, F. Moussa, and M. Spiesser, EPL. \textbf{29}, 251 (1995).
\bibitem{McGuire} M. A. McGuire, H. Dixit, V. R. Cooper, and B. C. Sales, Chem. Mater. \textbf{27}, 612 (2015).
\bibitem{Lebegue} S. Leb\'{e}gue, T. Bj\"{o}rkman, M. Klintenberg, R. M. Nieminen, and O. Eriksson, Phys. Rev. X \textbf{3}, 031002 (2013).
\bibitem{Li} X. Li and J. Yang, J. Mater. Chem. C \textbf{2}, 7071 (2014).
\bibitem{Lin} M. Lin, H. Zhuang, J. Yan, T. Z. Ward, A. A. Puretzky, C. M. Rouleau, Z. Gai, L. Liang, V. Meunier, B. G. Sumpter, P. Ganesh, P. R. C. Kent, D. B. Geohegan, D. G. Mandrus, and K. Xiao, J. Mater. Chem. C \textbf{4}, 315 (2016).
\bibitem{YangD} D. Yang, W. Yao, Q. Chen, K. Peng, P. Jiang, X. Lu, C. Uher, G. Wang, and X. Zhou, Chem. Mater. \textbf{28}, 1611 (2016).
\bibitem{Tang} X. Tang, D. Fan, K. Peng, D. Yang, L. Guo, X. Lu, J. Dai, G. Wang, H. Liu, and X. Zhou, Chem. Mater. \textbf{29}, 7401 (2017).
\bibitem{Tang1} X. Tang, D. Fan, L. Guo, H. Tan, S. Wang, X. Lu, X. Cao, G. Wang, and X. Zhou, Appl. Phys. Lett. \textbf{113}, 263902 (2018).
\bibitem{Peng} C. Peng, G. Zhang, C. Wang, Y. Yan, H. Zheng, Y. Wang, M. Hu, Phys. Status Solidi RRL \textbf{12} 1800172, (2018).
\bibitem{Williams} T. J. Williams, A. A. Aczel, M. D. Lumsden, S. E. Nagler, M. B. Stone, J. Q. Yan, and D. Mandrus, Phys. Rev. B \textbf{92}, 144404 (2015).
\bibitem{Vincent} H. Vincent, D. Leroux, D. Bijaoui, R. Rimet, and C. Schlenker, J. Solid State Chem. \textbf{63}, 349 (1986).
\bibitem{Rimet} R. Rimet, C. Schlenker, and H. Vincent, J. Magn. Magn. Mater. \textbf{25}, 7 (1981).
\bibitem{MAY} A. F. May, Y. Liu, S. Calder, D. S. Parker, T. Pandey, E. Cakmak, H. Cao, J. Yan, and M. A. McGuire, Phys. Rev. B \textbf{95}, 174440 (2017).
\bibitem{Martinez} L. M. Martinez, H. Iturriaga, R. Olmos, L. Shao, Y. Liu, T. T. Mai, C. Petrovic, A. R. H. Walker, and S. R. Singamaneni, Appl. Phys. Lett. \textbf{116}, 172404 (2020).
\bibitem{Sao} R. Olmos, J. A. Delgado, H. Iturriaga, L. M. Martinez, C. L. Saiz, L. Shao, Y. Liu, C. Petrovic, and S. R. Singamaneni, arXiv:2104.06564 (2021).
\bibitem{May} A. F. May, H. Cao, and S. Calder, J. Mag. Magn. Mater. \textbf{511}, 166936 (2020).
\bibitem{MB} M. B. Salamon and M. M. Jaime, Rev. Mod. Phys. \textbf{73}, 583 (2001).
\bibitem{YT} Y. Tokura, Rep. Prog. Phys. \textbf{69}, 797 (2006).
\bibitem{PRL} P. Majumdar and P. Littlewood, Phys. Rev. Lett. \textbf{81}, 1314 (1998).
\bibitem{Vineis} Christopher J. Vineis, Ali Shakouri, Arun Majumdar and Mercouri G. Kanatzidis, Adv. Mater. \textbf{22}, 3970 (2010).
\bibitem{Snyder} G. Jeffrey Snyder and Eric S. Toberer, Nature Materials \textbf{7}, 105 (2008).
\bibitem{Beekman} Matt Beekman, Donald T. Morelli and George S. Nolas, Nature Materials \textbf{14}, 1182 (2015).
\bibitem{ZhaoL} Li-Dong Zhao, Shih-Han Lo, Yongsheng Zhang, Hui Sun, Gangjian Tian, Citrad Uher, C. Wolverton, Vinayak P. Dravid and Mercouri G. Kanatzidis, Nature \textbf{508}, 373 (2014).
\bibitem{TakabatakeT} Toshiro Takabatake, Koichiro Suekuni, Tsuneyoshi Nakayama and Eiji Kaneshita, Rev. Mod. Phys. \textbf{86}, 669 (2014).
\bibitem{YuLiu} Y. Liu and C. Petrovic, Phys. Rev. B \textbf{98}, 064423 (2018).
\bibitem{PyFAI1} J. Kieffer and D. Karkoulis, PyFAI, a versatile library for azimuthal regrouping, J. Phys.: Conf. Ser. \textbf{425} 202012 (2013).
\bibitem{PyFAI2} J. Kieffer and J. P. Wright, PyFAI: a Python library for high performance azimuthal integration on GPU, Powder Diffraction \textbf{28} S339 (2013).
\bibitem{Toby} B. H. Toby and R. B. Von Dreele, GSAS-II: the genesis of a modern open-source all purpose crystallography software package, J. Appl. Cryst. \textbf{46}, 544 (2013).
\bibitem{Subias} G. Sub\'{\i}as, J. Garc\'{\i}a, M. G. Proietti, and J. Blasco, Phys. Rev. B \textbf{56}, 8183 (1997).
\bibitem{Prins} R. Prins and D. C.Koningsberger (eds.), X-ray Absorption: Principles, Applications, Techniques of EXAFS, SEXAFS, XANES (Wiley, New York, 1988).
\bibitem{Arrott1} A. Arrott, Phys. Rev. B \textbf{108}, 1394 (1957).
\bibitem{Austin} I. G. Austin, and N. F. Mott, Adv. Phys. \textbf{50}, 757 (2001).
\bibitem{Gopal} E. S. R. Gopal, Specific Heats at Low Temperatures (Plenum Press, New York, 1966).
\bibitem{HuangC} C. Y. Huang, J. Magn. Magn. Mater, \textbf{51}, 1 (1985).
\bibitem{Ni} Y. Ni, H. Zhao, Y. Zhang, B. Hu, I. Kimchi, and G. Cao, Phys. Rev. B \textbf{103}, L161105 (2021).
\end{references}
\end{document}